\documentclass[
showpacs,
pra,
twocolumn,
superscriptaddress,
aps,
amsmath,
amssymb
]{revtex4-2}

\usepackage[T1]{fontenc}
\usepackage[utf8]{inputenc}
\usepackage{newtxtext}
\usepackage{newtxmath}
\usepackage{microtype}
\usepackage{titlesec}
\usepackage{amsmath,amssymb,amsfonts}
\usepackage{mathtools}
\usepackage{bm}
\usepackage{physics}
\usepackage{bbold}

\usepackage{graphicx}
\graphicspath{{Figures/}}

\usepackage{float}
\usepackage{placeins}
\usepackage{multirow}
\usepackage{dcolumn}
\usepackage{booktabs}
\usepackage[caption=false]{subfig}
\usepackage{tikz}

\usepackage{xcolor}
\usepackage{pifont}
\usepackage{eurosym}

\usepackage[
breaklinks=true,
colorlinks=true,
citecolor=blue,
linkcolor=blue,
urlcolor=blue
]{hyperref}

\makeatletter
\renewcommand\@fnsymbol[1]{*}
\makeatother

\DeclareMathAlphabet{\bi}{OML}{cmm}{b}{it}

\newcommand{\be}{\begin{equation}}
\newcommand{\ee}{\end{equation}}

\titlespacing*{\section}{0pt}{3.5ex plus 1ex minus .2ex}{1.5ex plus .2ex}
\begin{document}
\title{Valley- and Orbital-Controlled 2D Chern Insulators Without Spin–Orbit Interaction}
\author{J. Benkaida}
\affiliation{LPHE, Modeling and Simulations, Faculty of Science, Mohammed V University in Rabat, Rabat, Morocco}
\affiliation{CPM, Centre of Physics and Mathematics, Faculty of Science, Mohammed V University in Rabat, Rabat, Morocco}
\author{O. Benhaida}
\affiliation{LPHE, Modeling and Simulations, Faculty of Science, Mohammed V University in Rabat, Rabat, Morocco}
\affiliation{CPM, Centre of Physics and Mathematics, Faculty of Science, Mohammed V University in Rabat, Rabat, Morocco}
\author{ L. B. Drissi}
\email[Corresponding author: ]{lalla-btissam.drissi@fsr.um5.ac.ma} 
\affiliation{LPHE, Modeling and Simulations, Faculty of Science, Mohammed V University in Rabat, Rabat, Morocco}
\affiliation{CPM, Centre of Physics and Mathematics, Faculty of Science, Mohammed V University in Rabat, Rabat, Morocco}
\affiliation{College of Physical and Chemical Sciences, Hassan II Academy of Sciences and Technology, Rabat, Morocco}
\author{ E. H. Saidi}
\affiliation{LPHE, Modeling and Simulations, Faculty of Science, Mohammed V University in Rabat, Rabat, Morocco}
\affiliation{CPM, Centre of Physics and Mathematics, Faculty of Science, Mohammed V University in Rabat, Rabat, Morocco}
\affiliation{College of Physical and Chemical Sciences, Hassan II Academy of Sciences and Technology, Rabat, Morocco}

\begin{abstract}
We present a theoretical study of orbital-induced topological phase transitions in a two-dimensional lattice model with staggered potential $(\Delta)$ and orbital coupling $(\lambda)$ competing with the hopping strength. By tuning these parameters, two gap-closing mechanisms emerge: valley closure at $\mathbf{K}$ and $\mathbf{K'}$ for $\lambda=\pm\Delta$, and a $\mathbf{\Gamma}$-point closure at $\lambda=\pm\sqrt{\Delta^2+9t_0^{2}}$. Their interplay defines a topological window in which the Berry curvature localizes near a single valley, yielding a quantized anomalous Hall conductivity ($\sigma_{xy}=e^{2}/h$) and Chern number ($C=1$). These results demonstrate orbital-driven Chern insulating behavior without spin–orbit coupling. The resulting phase diagram captures the transition from trivial to topological phases and suggests practical routes for orbital engineering in tunable lattice systems.\newline
		
	\end{abstract}
	\maketitle
\section{Introduction}

Over the past decade, advances in the study of topological phases have profoundly reshaped the understanding of quantum materials. The discovery of topological insulators \cite{Bernevig2006,Hasan2010} and related phenomena such as the Quantum Spin Hall Effect (QSHE) \cite{Fu2007}, marked a paradigm shift in condensed matter physics. These developments established the foundation for a broad class of materials whose exotic edge and surface states arise from topological invariants rather than conventional symmetry breaking. Topological systems characterized by integer Chern numbers exhibit rich and complex band structures that host multiple protected edge states \cite{Shafiei2023}. Such features open new frontiers for designing fault-tolerant quantum devices and low-dissipation electronic components. The Quantum Anomalous Hall Effect (QAHE), first predicted by Haldane and later realized experimentally \cite{Haldane1988, Rui2010, Zhong2017}, stands as a milestone example; demonstrating quantized Hall conductance without an external magnetic field.

In parallel, a great deal of effort has expanded toward two-dimensional lattices beyond graphene. A growing class of hexagonal and group IV derived monolayers; such as silicene, germanene, stanene, and plumbene, exhibit nontrivial topological properties driven by intrinsic or enhanced  spin–orbit coupling \cite{Liu2011, Xu2013, Ezawa2015}. These buckled honeycomb lattices, unlike planar graphene, enable electrical control of topological phase transitions through an applied perpendicular electric field, allowing the system to switch between a trivial insulator and a quantum spin Hall (QSH) state. First-principles studies have also revealed that related group-IV and group-V monolayers, including SiSn, PbBi, and BiSb, possess large nontrivial band gaps, making them promising candidates for room-temperature topological insulators  \cite{Cong2020, Bentaibi2024PbBi}.  Beyond elemental systems, heterostructures and functionalized graphene-based nanosheets, such as 2D OsC and graphyne derivatives, have emerged as tunable platforms for realizing topological phases with high Chern numbers \cite{Bosnar2023, Bentaibi2022OsC}. The incorporation of transition-metal atoms into graphene or hexagonal carbon frameworks enhances spin–orbit coupling and magnetic exchange, giving rise to exotic phenomena including the quantum anomalous Hall (QAHE) effect and topological magnetic semimetal phases \cite{Qiao2010,Zou2019}. Recent experimental and theoretical studies on phosphorene and other buckeled monolayers have further demonstrated how structural anisotropy and strain engineering can induce topological band inversions \cite{ Chen2020, Liu2018}.

More recently, the exploration of higher-order topological phases (HOTPs) has greatly broadened the classification of quantum matter beyond conventional edge and surface topologies \cite{Drissi2020}. In these systems, topological protection emerges in lower-dimensional boundaries, such as hinge or corner states, arising from quantized bulk multipole moments and crystalline symmetry constraints \cite{Benalcazar2017,Schindler2018, Drissi2021}. Some theoretical and computational advances have identified a rich variety of HOTPs in hexagonal and layered materials, including bismuth-based nanofilms, transition-metal dichalcogenides, and graphene-derived lattices \cite{Park2019,Drissi2022}. Experimental progress is also accelerating: scanning tunneling spectroscopy and transport measurements have confirmed quantized corner modes in fabricated 2D and 3D nanostructures, including bismuth, SnTe, and Kagome lattices \cite{Noguchi2021,Xue2019}. These breakthroughs reveal that higher-order topology is an experimentally accessible regime with immense potential for robust quantum-state engineering, nanoscale circuitry, and next-generation spintronic devices.

Recent advances have revealed that the orbital degree of freedom (ODoF) can serve as an emergent, spin-independent route to realizing and controlling topological phases and quantized transport phenomena. Unlike traditional mechanisms based on spin–orbit coupling, ODoF-based topology originates from inter-orbital hybridization and the geometric phase structure inherent to multi-orbital lattices \cite{Benkaida2025}. Experimental demonstrations in engineered wave systems have shown that orbital-selective hopping between orthogonal $p_x$ and $p_y$ orbitals—achieved through lattice geometry rather than on-site mixing—can produce robust, disorder-resilient edge states and even lasing in topological edge channels, proving the feasibility of orbital topology in the complete absence of intra-site orbital mixing \cite{St-Jean2017,Gao2023}. Similar phenomena have been identified in electronic and acoustic platforms, where controlled orbital hybridization yields quantized conductance and chiral edge transport, establishing orbital coupling as a new design paradigm for topological materials.

Beyond these early demonstrations, recent theoretical frameworks have greatly expanded our understanding of orbital-driven topology. It has been shown that inter-orbital hopping on two-dimensional lattices can generate orbital Hall conductance and non-trivial Chern numbers even in spinless systems \cite{Sun2020,Barbosa2024}. Moreover, orbital Chern insulators have been theoretically predicted and experimentally supported in moiré superlattices and low-symmetry 2D crystals, where orbital-projected band inversions induce quantized Hall responses even in the weak spin–orbit limit \cite{Fan2023}. The topological nature and Berry curvature structure of such orbital Chern states have been rigorously formalized only recently, providing a unified description of their quantum geometry and identifying material candidates such as blue-phosphorene and buckled honeycomb lattices \cite{Yao2026}. Complementarily, strong orbital magnetoelectric coupling has been observed in three-dimensional Chern insulators, confirming that orbital physics can mediate macroscopic electromagnetic responses \cite{Lu2024}. Within this broader context, the applied electric field emerges as a highly effective external control knob for tuning orbital hybridization and topological order. In buckled honeycomb lattices such as silicene, germanene, and their orbital analogues, the vertical displacement between sublattices induces a field-dependent staggered on-site potential that breaks inversion symmetry and lifts valley degeneracy \cite{Ezawa2012,Won2020}. This tunable potential enables continuous and reversible control over the band topology, driving transitions between quantum anomalous Hall, valley-polarized, and trivial insulating phases. 

In this study, we investigate the topological properties of a $p$-orbital buckled honeycomb lattice under a controllable electric field, with the aim of revealing how orbital and electrostatic degrees of freedom cooperatively determine the system’s phase behavior. Building upon recent progress in orbital topology and motivated by the pre-cited insights, we present a systematic analysis of the interplay between orbital coupling (OC), electric-field–induced sublattice potential, and their combined impact on the band inversion and Chern topology. Our approach demonstrates that tuning the orbital coupling parameter $\lambda$ modifies the effective inter-orbital hopping and drives nontrivial band inversions, while varying the electric potential $\Delta$ allows fine control over valley splitting and topological gaps. This dual tunability gives rise to a versatile phase diagram rich in topological transitions, including orbital Chern insulators, valley-polarized quantum anomalous Hall phases, and trivial insulators. As illustrated in Fig.~\ref{fig5}, the resulting phase diagram—parameterized by the orbital coupling strength $\lambda$ and the field-induced sublattice potential $\Delta$—captures the competition between inversion-symmetry breaking and orbital rehybridization. The present results establish orbital coupling and electric control as complementary mechanisms for designing and switching topological states in realistic two-dimensional systems, providing a pathway toward next-generation orbitronic and electro-topological devices.

\section{Tight binding model}

In the present work, we develop a multi-orbital tight-binding model based on the Slater–Koster formalism~\cite{Slater1954} to investigate the electronic structure and topological properties of a buckled honeycomb lattice.  
The Slater–Koster method provides a systematic framework for describing orbital couplings and their dependence on bond geometry.  
In this framework, the directional dependence of orbital overlap is captured through the directional cosines of the bond angles, which are crucial for evaluating the hopping integrals. 
This approach enables a clear understanding of the complex hopping dynamics inherent to the buckled honeycomb geometry, with particular emphasis on the ($p_x,p_y$) orbitals that dominate the low-energy physics.  The $p_z$ orbital is neglected since it is effectively passivated and energetically far from the Fermi level~\cite{Reis2017,Song2018}.  
As demonstrated by Reis \emph{et al.}~\cite{Reis2017}, $p_z$ orbitals typically participate in $\sigma$-bonding only when covalently functionalized, which pushes their energies well away from the Fermi level.  
This assumption is further supported by studies on chemically modified and substrate-supported systems, such as engineered honeycomb lattices on Cu(111) surfaces~\cite{Gardenier2020} and functionalized two-dimensional monolayers~\cite{Wan2023}, where $p_z$ states are effectively quenched.

Having reduced the orbital manifold to the in-plane ($p_x,p_y$) sector, we next determine the symmetry-allowed hopping amplitudes between neighboring orbitals.  
Within the Slater–Koster formalism, these amplitudes depend explicitly on the bond orientation through the two-center integrals $V_{pp\sigma}$ and $V_{pp\pi}$.  
For a bond direction $\hat{\mathbf{R}}=(n_{x},n_{y},n_{z})$, normalized such that $n_{x}^2+n_{y}^2+n_{z}^2=1$. Our polar angle $\theta$ is defined as the elevation measured from the $xy$-plane, so that $\theta = 0$ corresponds to an in-plane bond. This is related to the conventional spherical polar angle $\theta_{sph}$ (measured from the$z$ axis) by $\theta = \pi/2 -\theta_{sph}$, the direction cosines can be expressed as $n_{x,j}=\cos(\theta)\cos(\phi_j)$, $n_{y,j}=\cos(\theta)\sin(\phi_j)$, and $n_{z,j}=\sin(\theta)$.  
The hopping element between two orbitals $\alpha,\beta\in\{p_x,p_y\}$ then reads:  
\begin{equation}
t_{\alpha,\beta,i,j} = \langle p_{\alpha}^{i}|H_{0}|p_{\beta}^{j}\rangle 
= V_{pp\sigma}n_i n_j + (1-n_i n_j)V_{pp\pi}.
\end{equation}

\begin{table}[tbh]
	\centering
	\begin{tabular}{|c|c|}
		\hline
		Hopping element & Slater–Koster parameters \\ \hline
		$t_{p_xp_x}$ & $n_x^2V_{pp\sigma}+(1-n_x^2)V_{pp\pi}$ \\ 
		$t_{p_yp_y}$ & $n_y^2V_{pp\sigma}+(1-n_y^2)V_{pp\pi}$ \\ 
		$t_{p_xp_y}$ & $n_xn_y(V_{pp\sigma}-V_{pp\pi})$ \\ \hline
	\end{tabular}
	\caption{Slater–Koster hopping elements between neighboring sites. 
		Here $(n_{x},n_{y},n_{z})$ are the directional cosines of the bond connecting two neighboring atoms with respect to the $x$, $y$, and $z$ axes.}
	\label{Tsk}
\end{table}

\begin{figure*}[!t]
	\centering
	\includegraphics[width=0.8\textwidth]{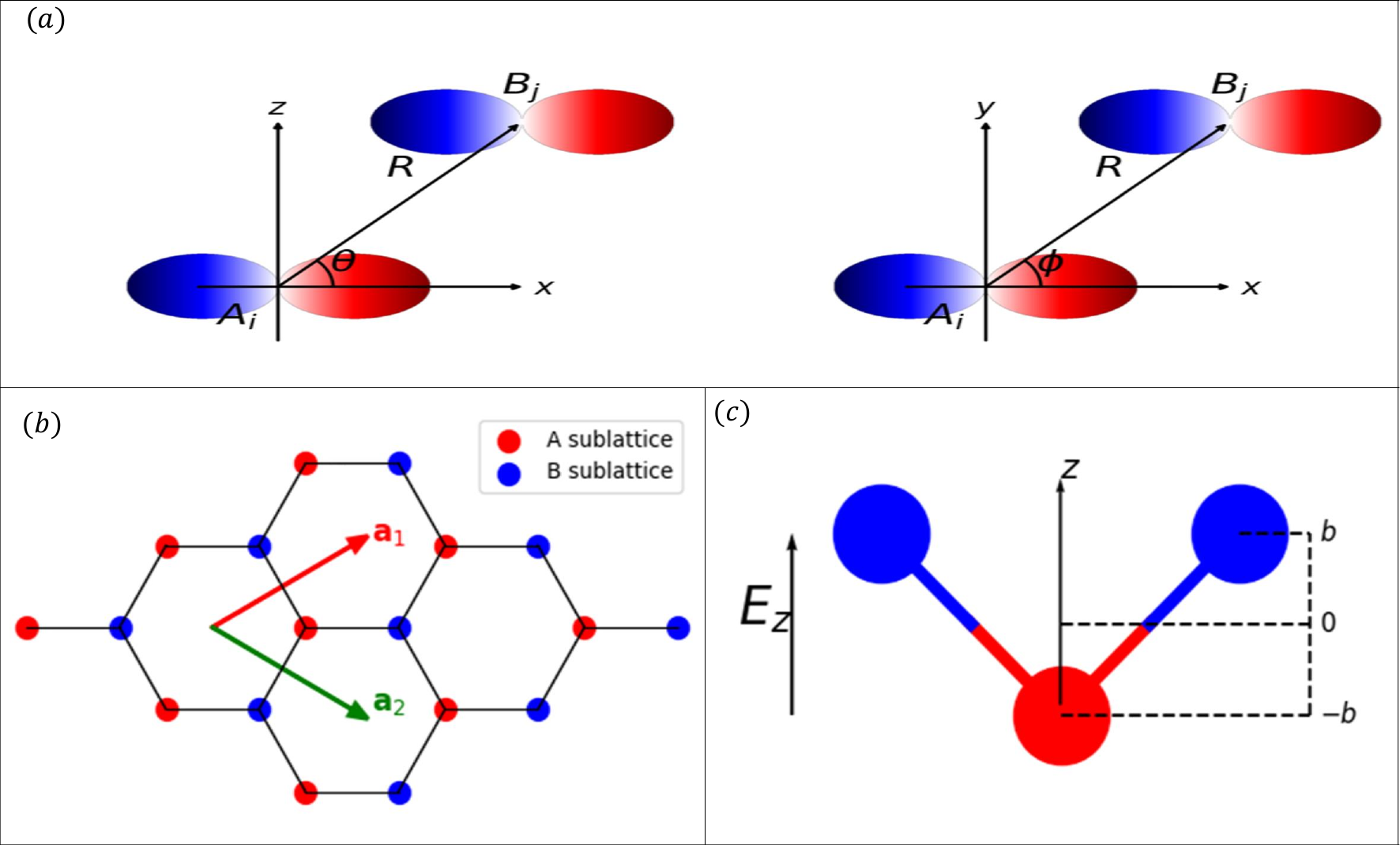}
	\caption{Lattice structure and orbital configuration.
			(a) Spatial configuration of $p_x$ orbitals. The orbital centered at the origin corresponds to sublattice $A$, while the second orbital belongs to sublattice $B$ at coordinates $(R, \theta, \phi)$.
			(b) Top view of the honeycomb lattice. The primitive lattice vectors for sublattices $A$ and $B$ are given by $a_{1,2}=(\frac{3}{2}a, \pm \frac{\sqrt{3}}{2}a, \mp b)$.
			(c) Side view showing sublattice $A$ and its nearest-neighbor $B$ sites. An applied vertical electric field $E_z$ generates a staggered potential  $\Delta = eE_z b$.}
	\label{fig1}
\end{figure*}

Figure~\ref{fig1}(a) schematically illustrates the spatial orientation and overlap of $p_x$ orbitals in the buckled lattice, where $\phi_j$ denotes the in-plane angle between the A–B bond and the $x$ axis.

Interactions between $p_x$ and $p_y$ orbitals can arise from multiple physical mechanisms, including lattice distortions, substrate-induced asymmetries, or applied external fields. 
These effects partially lift the orthogonality of the orbitals, giving rise to a finite orbital coupling $\lambda$.  
In cold-atom optical lattices, for instance, local distortions or anharmonic potential variations naturally induce such on-site orbital mixing, offering an ideal platform to simulate orbital hybridization phenomena.
It has been demonstrated that even small hybridization strengths can substantially influence the quantum phase behavior of $p$-band systems~\cite{Benkaida2025,St-Jean2017,Sun2020}.  

We formulate the system using a multi-orbital tight-binding Hamiltonian on a bipartite (A–B) lattice in the absence of an external field:  
\begin{equation}
\begin{split}
H_0 =& \sum_{i\alpha}\frac{\varepsilon_{i\alpha}}{2}(a_{i\alpha}^\dagger a_{i\alpha}+b_{i\alpha}^\dagger b_{i\alpha})+ \sum_{\langle i\alpha,j\beta\rangle} t_{i\alpha,j\beta} a_{i\alpha}^\dagger b_{j\beta}\\
&- i\lambda \sum_{j,\alpha\neq\beta}(a_{j,\alpha}^\dagger a_{j,\beta}+b_{j,\alpha}^\dagger b_{j,\beta}) + h.c.
\end{split}
\end{equation}
The first term represents the orbital-dependent on-site energies $\varepsilon_{i\alpha}$, identical for both sublattices.  
The second term accounts for nearest-neighbor hopping within the LCAO framework~\cite{Slater1954}, and the third term introduces the local intra-sublattice orbital coupling $\lambda$.

Unlike planar graphene, the buckled honeycomb lattices (e.g., silicene, germanene) possess an intrinsic out-of-plane displacement $b$ between sublattices A and B.
An external electric field $\vec{\mathbf{E}} = E_z\hat{z}$ couples to this vertical asymmetry, inducing a staggered potential that acts as a mass term:  
\begin{equation}
H_E = \Delta \sum_{j,\alpha}(a_{j,\alpha}^\dagger a_{j,\alpha} - b_{j,\alpha}^\dagger b_{j,\alpha}),
\end{equation}
where $\Delta = eE_z b$ explicitly breaks inversion symmetry and opens a tunable gap at the Dirac points.  
As illustrated in Fig.~\ref{fig1}(b)-(c), sublattice A atoms reside at $z_A=-b$, while sublattice B atoms are elevated to $z_B=+b$.

By applying a spatial Fourier transform to the field operators, defined as $c_{j,\alpha} = \frac{1}{\sqrt{N}} \sum_{\mathbf{k}} e^{i\mathbf{k}\cdot\mathbf{R}_j} c_{\mathbf{k},\alpha}$ (and similarly for the creation operators), the total Hamiltonian $H = H_0 + H_E$ can be expressed in momentum space as: $H = \sum_{\mathbf{k}} \psi_{\mathbf{k}}^\dagger \mathcal{H}(\bm k, \lambda, \Delta) \psi_{\mathbf{k}}$. In this expression, the four-component spinor is defined in the basis of the $p$-orbitals as $\psi_{\bm k} = (a_{\bm k,p_x}, a_{\bm k,p_y}, b_{\bm k,p_x}, b_{\bm k,p_y})^T$. The corresponding Bloch Hamiltonian $\mathcal{H}(\bm k, \lambda, \Delta)$ is given by:
\textbf{
	\begin{equation}
	H(\bm k, \lambda, \Delta)=
	\begin{pmatrix}
	h_{c}(\Delta, \lambda)& h(\bm k)\\
	h^{\dagger}(\bm k) & -h^{T}_{c}(\Delta, \lambda) \\
	\end{pmatrix},
	\end{equation}
	and 
	\begin{equation}
	h_{c}(\Delta, \lambda)=
	\begin{pmatrix}
	\Delta & -i\lambda \\
	i\lambda & \Delta 
	\end{pmatrix},\quad h(\bm k)=
	\begin{pmatrix}
	f_{p_xp_x}(\bm k) & f_{p_xp_y}(\bm k) \\
	f_{p_yp_x}(\bm kk) & f_{p_yp_y}(\bm k) 
	\end{pmatrix},
	\end{equation}}
with $f_{\alpha\beta}(\bm k)=\sum_{l=1}^{3}t_{l,\alpha\beta}e^{i\mathbf{k}\cdot\boldsymbol{\delta}_l}$ 
the Fourier-transformed hopping amplitudes between orbitals $\alpha$ and $\beta$ across sublattices.  

In this buckled geometry, sublattice A atoms form a plane parallel but vertically displaced relative to sublattice B, as shown in Fig.~\ref{fig1}(c). 
Time-reversal symmetry (TRS) in the spinless case corresponds to complex conjugation, $\mathcal{T}=\mathcal{K}$.  
While $H(k,0,\Delta)$ satisfies $\mathcal{T}H(k,0,\Delta)\mathcal{T}^{-1}=H(-k,0,\Delta)$,  
the orbital coupling term is odd under $\mathcal{K}$, leading to $\mathcal{T}H(k,\lambda,\Delta)\mathcal{T}^{-1}\neq H(-k,\lambda,\Delta)$.  
Hence, the finite orbital coupling $\lambda$ explicitly breaks TRS, a prerequisite for the emergence of nonzero Berry curvature and topologically nontrivial phases. For $\Delta=0$, the Hamiltonian $H(k,\lambda,0)$ preserves particle–hole symmetry (PHS) under $\mathcal{C}=U\mathcal{K}$, with $U=\tau_z\otimes I_2$.  
However, introducing $\Delta\neq0$ breaks this symmetry since the staggered potential is not invariant under $\mathcal{C}$. Moreover, the staggered potential $\Delta$ explicitly breaks the inversion symmetry. Under the inversion symmetry operator $\mathcal{P}=\tau_x\otimes I_2$, which exchanges the sublttices $A$ and $B$, $ \mathcal{P}H(k,\lambda,\Delta)\mathcal{P}^{-1}\neq H(-k,\lambda,\Delta)$. Consequently both orbital coupling and staggered potential break TRS and PHS respectively,also the Hamiltonian  $H(k,\lambda,\Delta)$ possesses no chiral symmetry. The system therefore belongs to Altland–Zirnbauer Class A, characterized in two dimensions by a Chern topological invariant~\cite{Altland1997,Ryu2010}:  
\begin{equation}
\mathcal{C} = \frac{1}{2\pi}\int_{\text{BZ}}\Omega(\mathbf{k})\,d^2\mathbf{k},
\end{equation}
where $\Omega(\mathbf{k})$ is the total Berry curvature of the occupied bands~\cite{Haldane1988}.
\section{Results and Discussion}
\subsection{Band structure}
\begin{figure*}[!t]
	\centering
	\includegraphics[width=1\textwidth]{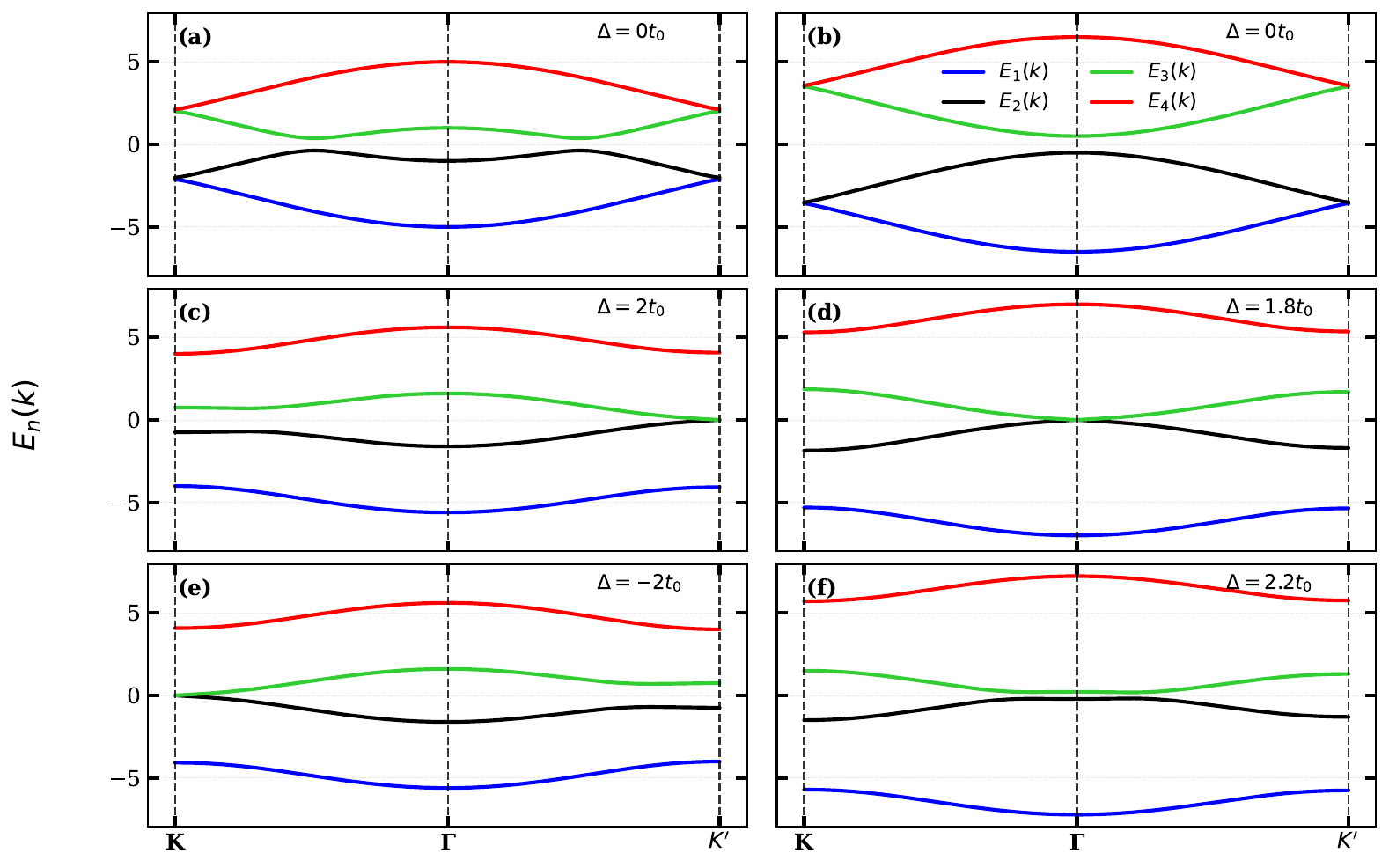}
	\caption{Energy band structures along the high-symmetry path ($\mathbf{K}–\mathbf{\Gamma}–\mathbf{K'}$) for various values of the orbital coupling $\lambda$ and the staggered potential $\Delta$. Panels (a), (c), and (e) correspond to $\lambda = 2t_0$, while panels (b), (d), and (f) correspond to $\lambda = 3.5t_0$. The value of the staggered potential $\Delta$ is indicated within each panel. The hopping parameter is fixed at $t_1 = 0.25t_0$.}
	\label{fig2}
\end{figure*}

The central narrative of our work is the competition betwenn two energy scales: the on-site orbital coupling $\lambda$ and the staggered sublattice potential $\Delta$. To understan this conpetition and its topological consequences, we must first establish the distinct roles of these parameters. The orbital coupling, $\lambda$, acting within the $\{p_{x},p_{y}\}$ subspace, is purely imaginary and takes the form $\pm i\lambda$.This operator is proportional to the orbital angular momentum operaopr $l_{z}$ and acts as an effective, intrinsic magnetic field that breaks the time-reversal symmetry. Its role is hybridize the in-plane orbitals, providing the band mixing necessary for the non-trivial phase. In stark contrast, the staggered potential, $\Delta$, induced by an external electric field, breaks the inversion symmetry. Its places sublattice $A$ and $B$ at different on-site energies, explicitly lifting any valley degeneracy amd acting as \textbf{valley selector}.
To investigate the influence of model parameters on the electronic spectrum, we calculate the band structure for various values of the on-site potential $\Delta$ and the orbital coupling strength $\lambda$, while keeping the hopping amplitude fixed at $t_1 = 0.25 t_0$. By analytically diagonalizing the Bloch Hamiltonian, we derive the exact expressions for the eigenvalues $E_{n}(\bm k,\lambda,\Delta)$ and their corresponding eigenvectors $|u_n(\bm k, \lambda, \Delta)\rangle$ for each band. These analytical solutions are expressed in the following form:
\begin{align}
E_1(\bm{k},\lambda,\Delta)= -\dfrac{1}{\sqrt{2}}\sqrt{\varPi(\bm{k},\lambda,\Delta)+\sqrt{\Lambda(\bm{k},\lambda,\Delta)}},
\end{align}
\begin{align}
E_2(\bm{k},\lambda,\Delta) = -\dfrac{1}{\sqrt{2}}\sqrt{\varPi(\bm{k},\lambda,\Delta)-\sqrt{\Lambda(\bm{k},\lambda,\Delta)}},
\end{align}
\begin{align}
E_3(\bm{k},\lambda,\Delta) = \dfrac{1}{\sqrt{2}}\sqrt{\varPi(\bm{k},\lambda,\Delta)-\sqrt{\Lambda(\bm{k},\lambda,\Delta)}}, 
\end{align}
\begin{align}
E_4(\bm{k},\lambda,\Delta) = \dfrac{1}{\sqrt{2}}\sqrt{\varPi(\bm{k},\lambda,\Delta)+\sqrt{\Lambda(\bm{k},\lambda,\Delta)}},
\end{align}
\begin{align}
|u_n(\bm{k}, \lambda, \Delta)\rangle &= \mathcal{N}_{n} \begin{bmatrix} \eta^{n}_{1}(\bm{k},\lambda,\Delta) & \eta^{n}_{2}(\bm{k},\lambda,\Delta) & 1 & \eta^{n}_{3}(\bm{k},\lambda,\Delta) \end{bmatrix}^{T},
\end{align}
and
	\begin{equation}
	\begin{split}
	\varPi(\bm k,\lambda,\Delta)
	=\;&
	2\Delta^2+2\lambda^2
	+\left|f_{p_{x}p_{x}}(\bm k)\right|^2
	 +2\left|f_{p_{x}p_{y}}(\bm k)\right|^2\\
	+&\left|f_{p_{y}p_{y}}(\bm k)\right|^2,
	\end{split}
	\end{equation}
	\begin{equation}
	\begin{aligned}
	\Lambda(\bm{k},\lambda,\Delta) = & -16\Delta\,\lambda\,\operatorname{Im}\! \left[ \bigl(f_{p_{x}p_{x}}(\bm{k}) - f_{p_{y}p_{y}}(\bm{k})\bigr) f^{*}_{p_{x}p_{y}}(\bm{k}) \right] \\
	& + \Big( |f_{p_{x}p_{x}}(\bm{k})|^2 - |f_{p_{y}p_{y}}(\bm{k})|^2 \Big)^2 \\
	& + 4|f_{p_{x}p_{y}}(\bm{k})|^2 \Big( |f_{p_{x}p_{x}}(\bm{k})|^2 + |f_{p_{y}p_{y}}(\bm{k})|^2 \Big) \\
	& + 8\operatorname{Re}\! \left[ f_{p_{x}p_{x}}(\bm{k}) f_{p_{y}p_{y}}(\bm{k}) f^{*\,2}_{p_{x}p_{y}}(\bm{k}) \right] \\
	& + 4\lambda^2 \Big( |f_{p_{x}p_{x}}(\bm{k})|^2 + |f_{p_{y}p_{y}}(\bm{k})|^2 \\
	& \quad + 2\operatorname{Re}\! \left[ f_{p_{x}p_{x}}(\bm{k}) f^{*}_{p_{y}p_{y}}(\bm{k}) \right] \Big) + 16\Delta^2\,\lambda^2,
	\end{aligned}
	\end{equation}

\begin{align}
\mathcal{N}_{n}(\bm{k},\lambda,\Delta) &= \left[ \sum_{i=1}^{3}\left|\eta^{n}_{i}(\bm{k},\lambda,\Delta)\right|^{2}+1\right]^{-1/2}, \\[1.5mm]
\eta^{n}_{1}(\bm{k},\lambda,\Delta) &= m^{n}_{11}(\bm{k},\lambda,\Delta)+m^{n}_{12}(\bm{k},\lambda,\Delta)\eta^{n}_{3}(\bm{k},\lambda,\Delta), \\
\eta^{n}_{2}(\bm{k},\lambda,\Delta) &= m^{n}_{21}(\bm{k},\lambda,\Delta)+m^{n}_{22}(\bm{k},\lambda,\Delta)\eta^{n}_{3}(\bm{k},\lambda,\Delta), \\
\eta^{n}_{3}(\bm{k},\lambda,\Delta) &= \dfrac{1}{z^{n}_{12}(\bm{k},\lambda,\Delta)}\big(1-z^{n}_{11}(\bm{k},\lambda,\Delta)\big).
\end{align}
\begin{align}
t_{1} &= \left[V_{pp\sigma }-V_{pp\pi }\right]\frac{\cos^{2}(\theta)}{2}, \\[1.5mm]
t_{0} &= \frac{1}{2}\left[V_{pp\sigma }\cos^{2}(\theta) +\big(2-\cos^{2}(\theta)\big)V_{pp\pi }\right],
\end{align}
	where $m^{n}_{ij}(\bm k,\lambda,\Delta)$ and $z^{n}_{ij}(\bm k,\lambda,\Delta)$ denote the matrix elements of the auxiliary matrices $\mathcal{M}^{n}(\bm k,\lambda,\Delta) = (E_{n}(\bm k,\lambda,\Delta) \mathbb{I} - h_{c}(\Delta, \lambda))^{-1} h(\bm k)$ and $\mathcal{Z}^{n}(\bm k,\lambda,\Delta) = (E_{n}(\bm k,\lambda,\Delta) \mathbb{I} + h^{t}_{c}(\Delta, \lambda))^{-1} h^{\dagger}(\bm k) \mathcal{M}$, respectively. Here, $\theta$ represents the angle between the A–B bond and the atomic plane of sublattice A. The system exhibits four energy bands: two valence bands ($n=1,2$) and two conduction bands ($n=3,4$). The calculated dispersion relations $E_n(\bm k)$ and the corresponding band structures are presented in Fig.~\ref{fig2}.
In the absence of an electric filed $(\Delta=0)$, the spectrum is gapped at the ($\mathbf{K}$) and  ($\mathbf{K'}$) valleys, as shown in fig.\ref{fig2}-(\textbf{a}) and \ref{fig2} -(\textbf{b}). When the finite ($\Delta$) is switched on, it explicitly breaks the inversion symmetry by shifting the on-site energies of sublattices A and B in oposite directions. This lifts the valley degeneracy: the band gap at one valley narrows while the gap at the other widens, depending on the sign of the ($\Delta$). In this regime the effect of the staggered potential has not yet compensated by the action of the orbital coupling, and the first topological phase transition occurs when the orbital hybridization induced by $\lambda$ exactly compensates the sublattice energy splitting. The condition for this valley-selective gap closure is given by $\lambda \pm \Delta = 0$, as illustrated in Fig. \ref{fig2}-(c) and \ref{fig2})-(e).
	At these critical points the conduction and valence bands touch at ($\mathbf{K}$) or ($\mathbf{K'}$), and their orbital character is inverted. This is a Valley-polarised band inversion: the swapping of band character occurs predominantly at one valley because of $\Delta$ has broken the symmetry between the two valleys, the reopening of the gap beyond this point sign as a change of topological invariant. A second, qualitatively different gap-closure mechanism appears at larger ($\lambda$). As shown in the fig \ref{fig2}-(\textbf{d}) and \ref{fig2}-(\textbf{f}), the gap closes at the Brillouin-zone centre ($\Gamma$) when ($\lambda\pm \sqrt{\Delta^2+9t_{0}^2}$).At this $\Gamma$-point transition the orbital character is again rearranged, but this time the process destroy the nontrivial topology and drives the system back into a normal insulating phase when  ($\lambda > \sqrt{\Delta^2+9t_{0}^2}$). The two critical lines $\lambda=\Delta$ and $\lambda= \sqrt{\Delta^2+9t_{0}^2}$ thus define a topological window, inside which the ground state is a Chern insulator with $C=1$. Outside this window the system is a trivial insulator $C=0$. The physical interpretation of eq (\ref{window}) is transparent: the $\lambda$ must be large enough to overcome the inversion-symmetry-breaking mass $\Delta$ and invert the valley band ordering, yet not so strong that it also close the gap at $\Gamma$ and re-trivialises the band structure\newline
\begin{equation}
|\Delta|<|\lambda|<|\sqrt{\Delta^2+9t_{0}^2}|
\label{window}.
\end{equation}
Both mechanisms play a decisive role in determining the electronic phase transitions. Similar valley-dependent gap closings have been reported in other multiorbital honeycomb systems such as the $\alpha$–T$_3$ lattice under light-induced perturbations \cite{Benhaida2025a,Tamang2024}, and in buckled silicene under external electric fields \cite{Ezawa2012}. Recent works on orbital Chern insulators \cite{Li2024} further confirm that interplay between interorbital coupling and sublattice potential leads to topological band inversions, consistent with our findings.
\begin{figure*}[!t]
	\centering
	\includegraphics[width=0.7\textwidth]{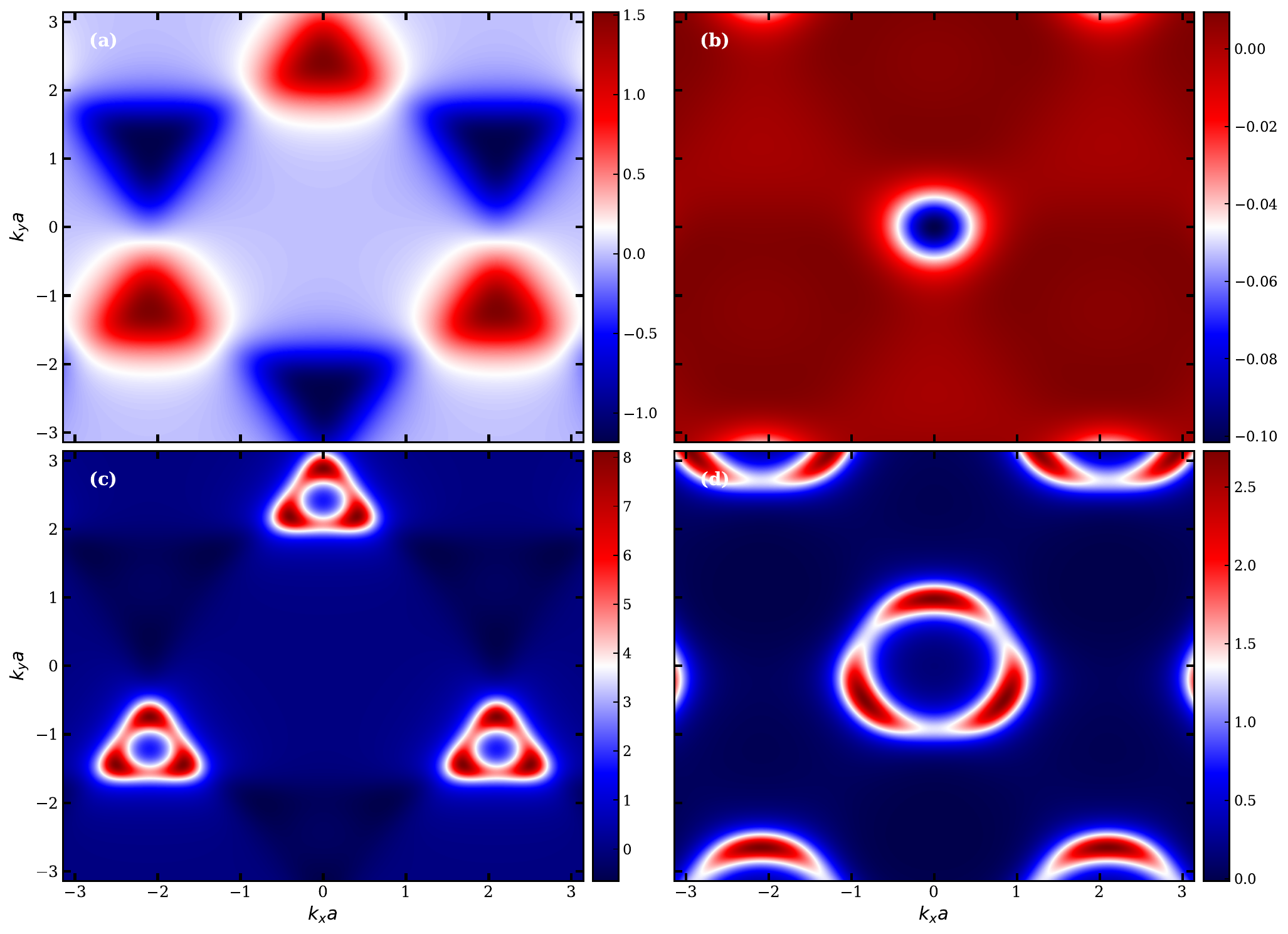}
	\caption{Distribution of the total Berry curvature of the occupied bands, $\Omega(\mathbf{k})/a^2$, in the momentum space ($k_x, k_y$) for various orbital coupling strengths $\lambda$. The panels show results for: (a) $\lambda = 0.5\, t_0$, (b) $\lambda = 4\, t_0$, (c) $\lambda = 1.2\, t_0$, and (d) $\lambda = 2.5\, t_0$. The hopping parameter is fixed at $t_1 = 0.25\, t_0$ and the staggered potential at $\Delta = t_0$.}
	\label{fig3}
\end{figure*}
\begin{figure*}[!t]
	\centering
	\includegraphics[width=1\textwidth]{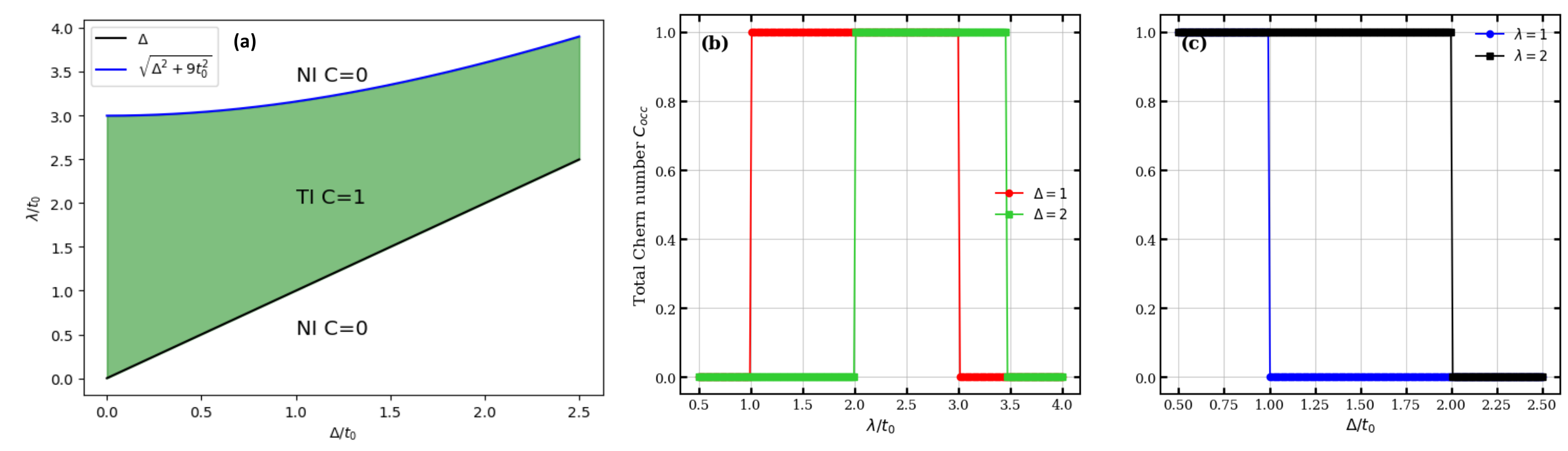}
	\caption{Variation in the total occupied bands' Chern number ($C_{occ} = C_1 + C_2$) regarding system parameters: (a) topological phase diagram in the ($\lambda, \Delta$) plane; (b) total Chern number dependence on the orbital coupling $\lambda$ for fixed $\Delta$ values (indicated within the panel); (c) total Chern number versus staggered potential $\Delta$ for fixed $\lambda$ values (indicated within the panel). The hopping parameter is fixed at $t_1 = 0.25 t_0$.}
	\label{fig5}
\end{figure*}
\subsection{Berry curvature and Chern number}
Having identified the critical points where the band gap closes, we now proceed to the topological characterization of the electronic phases.  
The Berry curvature $\Omega(\mathbf{k})$ \cite{Haldane1988,Benhaida2025a,Benhaida2025c} provides a direct measure of the local geometrical properties of the Bloch states in momentum space.  
Its integral over the Brillouin zone defines the Chern number, 
\[
C_{n} = \frac{1}{2\pi}\int_{\text{BZ}}\Omega_{n}(\bm k)\,d^{2}k,
\]
which serves as a topological invariant distinguishing different insulating phases.\newline
For an $n$th energy band, the Berry curvature $\Omega_n(\mathbf{k})$ can be computed from the Kubo formula~\cite{Thouless1982}:

	\begin{equation}
	\Omega_n(\mathbf{k}) = -2\,\hbar^{2}\text{Im}\sum_{m\neq n}
	\frac{
			M^{x}_{nm}
		M^{y}_{mn}
	}{
		(E_n(\bm k, \lambda, \Delta) - E_m(\bm k, \lambda, \Delta))^2
	},
	\end{equation}
	and
	\begin{equation}
	M^{i}_{nm}
	=
	\bra{u_n(\bm k, \lambda, \Delta)}
	v_i
	\ket{u_m(\bm k, \lambda, \Delta)},
	\end{equation}
	where $v_i=\hbar^{-1}\partial_{k_i}H(\bm k, \lambda, \Delta)$ represents the velocity component of the electron in the $i$-direction ($i \in \{x, y\}$) of the lattice plane.  The total Berry curvature of occupied bands is then $\Omega(\mathbf{k})=\sum_{n,E_n<0}\Omega_n(\mathbf{k})$.

The geometrical origine of the topological phase can be directly bisualised throught the total Berry cuvature of the occupied bands $\Omega(k)=\Omega_{1}(k)+\Omega_{2}(k)$. figurer \ref{fig3} shows $\Omega(k)$ for four charatistic values of $\lambda$ at fixed $\Delta=t_{0}$ and $t_{1}=0.25t_{0}$. In figure \ref{fig3}-(\textbf{a}) $\lambda<\Delta$ , the Berry curvature is diffuse, with broad pathes of opposite sign near the ($\mathbf{K}$) and ($\mathbf{K'}$) valleys, the total flux integral to exactly to zero, confirm a trivial phase with $C=0$. While in the figure fig \ref{fig3}-(\textbf{b}) $\lambda>\sqrt{\Delta^2+9t_{0}^2}$ , the berry curvature becomes very small and nearly uniform, again yielding $C=0$. The second gap closure at $\Gamma$ has removal all nontrivial Berry-phase contribution. in contrast Inside the topological window, fig \ref{fig3}-(\textbf{c}) and \ref{fig3}-(\textbf{d}), the Berry curvature collapses into sharp centred at ($\mathbf{K}$) or ($\mathbf{K'}$) points. The intensity is highly asymmetric: one valy carries a much larger peak than the other. This asymmetry is the momentum-space fingerprint of the vally polarisation caused by the sign of $\Delta$.\newline
	The physical reason for the strong localisation is direct consequence of the band inversion. the Berry curvature is inversely proportional to the band gap $E_{g}$. When the $\lambda$ drives a band inversion at one valley, the gap at that valley becames minimal, and the Berry curvature concentrates there. Integrating this localised flu over the whole Brilouin zone gives a net Chern $C=1$. In this sense, the orbital coupling $\lambda$ creates the magnetic field in momentum space. while the staggered potential $\Delta$ selects the valley where the flux is deposited. The underlying mechanism of the anomalous quantum Hall effect (AQHE) is the nonzero Berry curvature of the occupied bands~\cite{Haldane1988,Benhaida2025b}.  
In topological insulators, this curvature acts as an effective magnetic field in momentum space, producing a transverse Hall current.  
The intrinsic component of the AQHE, which we focus on here, is obtained from the Berry curvature via the Kubo formalism~\cite{Xiao2010,Jungwirth2002}: 
\textbf{
	\begin{equation}
	\sigma_{xy} = \frac{e^{2}}{\hbar}\sum_{n}\int_{\text{BZ}}\frac{d^{2}k}{(2\pi)^{2}}\,f_n(\mathbf{k})\,\Omega_n(\mathbf{k}),
	\end{equation}}
where $f_{n}(\mathbf{k})=\frac{1}{1+e^{\beta(E_{n}-\mu)}}$ is the Fermi–Dirac distribution.\newline 
\begin{figure}[h]
	\centering
	\includegraphics[width=0.5\textwidth]{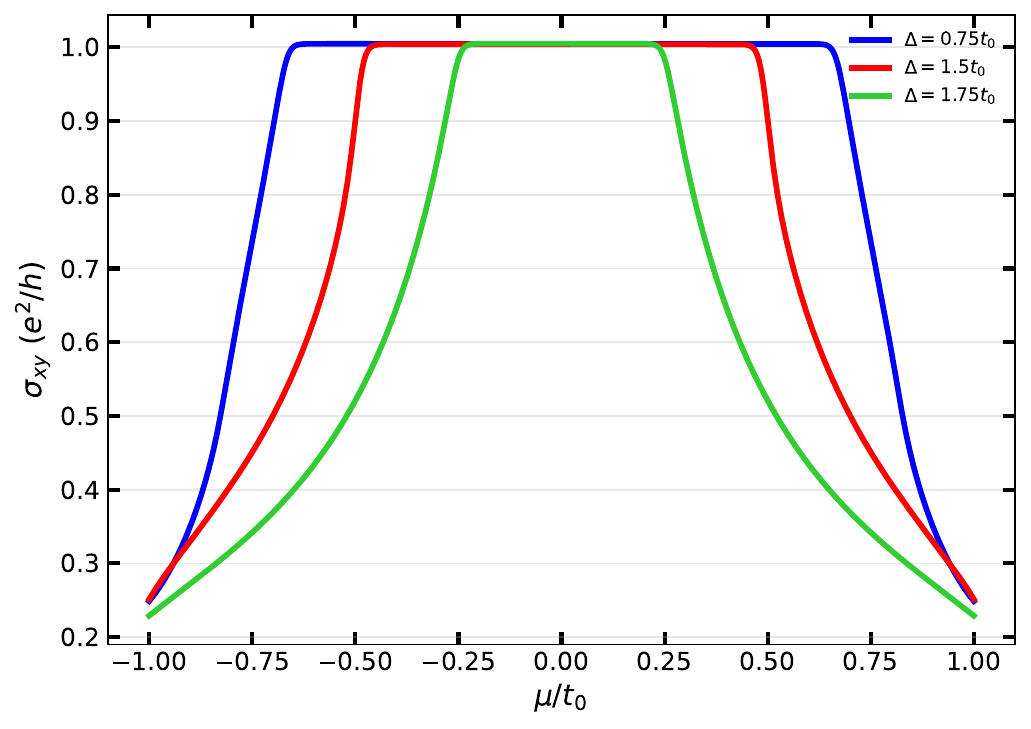}
	\caption{Anomalous Hall conductivity $\sigma_{xy}$ (in units of $e^2/h$) as a function of the chemical potential $\mu/t_0$ for various staggered potentials $\Delta$ (as labeled in the figure). The parameters are set to $t_1 = 0.5\, t_0$ for the hopping and $\lambda = 2\, t_0$ for the orbital coupling.}
	\label{fig4}
\end{figure}
 Figure \ref{fig4} shows $\sigma_{xy}(\mu)$ for several values of $\Delta$ with $\lambda=2t_{0}$. When  the chimical potential lies inside the bulk gap $\mu\approx 0$, a perfectly quantised plateau  $\sigma_{xy}=\frac{e^2}{h}$ appears, corrsponding to $C=1$. The plateau is intensitive to small change in $\mu$, a hallmark of topological protection. As $\mu$ enters the conduction or valence bands, the plateau vanishes and   $\sigma_{xy}$ decrease smoothly because bulk carriers begin to contribute.\newline
	Finally,Fig\ref{fig5}-\textbf{a} presents the complet topological phase diagram in the ($\Delta,\lambda$) plane. The colored region, bounded by the two gap-closing lines, is the Chern insualting phase with $C=1$.\newline
Figures~\ref{fig5}(b) and \ref{fig5}(c) summarize the dependence of the Chern number on $\lambda$ and $\Delta$, respectively.  
For a fixed $\Delta=t_{0}$, the Chern number changes from $C=0$ to $C=1$ at $\lambda_{c}=\Delta$, consistent with the valley-gap closure condition.  
For larger $\Delta=2t_{0}$, the transition shifts to $\lambda_{c}=2t_{0}$, and $C$ returns to zero when $\lambda > \sqrt{\Delta^{2}+9t_{0}^{2}}$.  
Conversely, when fixing $\lambda$ and varying $\Delta$, the transitions occur in reverse: for $\lambda=t_{0}$, $C$ changes from $1$ to $0$ at $\Delta_{c}=\lambda$, and similarly for $\lambda=2t_{0}$. Overall, the results show that the topological nature of the insulating phase is dictated by the relative magnitudes of $\lambda$ and $\Delta$. The system is a trivial insulator ($C=0$) where $|\lambda|<|\Delta|$ or $|\lambda|>\sqrt{\Delta^{2}+9t_{0}^{2}}$, and it's a Chern insulator ($C=1$) inside the window $\sqrt{\Delta^{2}+9t_{0}^{2}} > |\lambda| > |\Delta|$. This behavior mirrors the phase evolution found in other topological multi-orbital  HgTe honeycomb lattices \cite{Beugeling2015}, confirming that orbital coupling acts as an effective pseudo spin–orbit interaction capable of driving topological phase transitions even in spinless systems.\newline
 The model is directly relevant to two-dimensional materials where the $p_{z}$ orbital is removed from the low-ebergy sector by substrate or chemical functionalisation. For example,in stanene on Cu(111) and in bismuthene on Ag(111) the strong hybridisation of Bi-$p_{z}$ states with the substrate shifts these bands below the fermi level, a substrate-orbital-filtring effect confirmed by DFT calculations \cite{Sun2021,Deng2018}. Similarly, functinalised germanene exihibit fermi-surface dominated by in-palne $p_{x}$ and $p_{y}$ orbitals \cite{Ren2018,Qu2016}, and the choice of the hopping parameter ratio ($t_{1}/t_{0}$) used in our model is motivated by the recent work \cite{Benkaida2025}.Finaly when a realistic Slater-koster parameter are taken for Silicene, Germanene, and Stanene table\ref{Tab2} \cite{Hattori2017}, the resulting ratio ($t_{1}/t_{0}$) is generally larger.Neverthless, the topological phase of the system is is robust: For nonzero orbital coupling ($0<\lambda<3t_{0}$) , the ground stat is Chern insulator with ($C=1$) \cite{Benkaida2025} as long as the staggered potential statisfies $\Delta<\lambda$.
\begin{table}[tbh]
	\centering
	\caption{Numerical values of parametrs in multi-orbital tight-binding model are chosen according to Ref \cite{Hattori2017}}
	\begin{tabular}{|c|c|c|c|c|c|}\hline
		Systwm &$a(\AA)$&$\theta^\circ$&$V_{pp\sigma}$(eV)&$V_{pp\sigma}$(eV)&$t_{1}/t_{0}$\\\hline
		Silicene &3.86&11.7&4.47&-1.12&1.72\\  \hline
		Germanene &4.02&16.5&4.15&-1.04&1.78\\  \hline
		Stanene &4.70&17.1&1.49&-0.79&4.33\\  \hline
	\end{tabular}
	\label{Tab2}
\end{table}
\section{Conclusion}
We have theoretically investigated the emergence of topological phases in a two-dimensional lattice model where orbital degrees of freedom interact with a staggered potential and hopping asymmetry. By varying the orbital coupling strength $\lambda$ and the sublattice potential $\Delta$, we identified two distinct mechanisms of gap closure: a valley-driven transition at $\mathbf{K}$ and $\mathbf{K'}$ points when $\lambda=\pm\Delta$, and a central ($\mathbf{\Gamma}$)-point transition at $\lambda=\pm\sqrt{\Delta^{2}+9t_{0}^{2}}$. These critical lines separate topologically trivial and nontrivial insulating phases. Calculations of the Berry curvature and Chern number reveal a quantized plateau in the anomalous Hall conductivity ($\sigma_{xy}=e^{2}/h$) for intermediate coupling, confirming the existence of a Chern insulating phase. The phase diagram constructed in the $(\Delta,\lambda)$ plane shows that the topological character of the system is entirely governed by the relative magnitudes of these two parameters, providing a unified description of orbital-induced topological transitions.\newline
Our results demonstrate that even in spinless systems, orbital hybridization can mimic spin–orbit coupling, giving rise to nontrivial topology and quantized Hall transport. This establishes orbital degrees of freedom as a new and versatile platform for realizing Chern insulators without relying on magnetic order or external fields. Such control of topological states through orbital engineering could be directly relevant to optically driven lattices, and cold-atom systems. Future work may extend this approach to include interactions, disorder, and light–matter coupling to explore Floquet and correlated topological phenomena within the same orbital framework.
\section{Acknowledgment}
The authors would like to acknowledge the "Academie Hassan II des Sciences 
et Techniques"-Morocco for its financial support. The authors also thank the
LPHE-MS, Faculty of Sciences, Mohammed V University in Rabat, Morocco for 
the technical support through computer facilities, where all the 
calculations have been performed.\newline
\textbf{Conflict of Interest}: The authors declare that they have no known competing 
financial interests or personal relationships that could have appeared to 
influence the work reported in this paper.\newline
\textbf{Data availability statement}: The data that support the findings of this
study are available  upon reasonable request from the authors.


\newpage
\begin{thebibliography}{100}
	\bibitem[1]{Bernevig2006} B. A. Bernevig, T. L. Hughes, S. C. Zhang,
	\textit{Quantum Spin Hall Effect and Topological Phase Transition in HgTe Quantum Wells},
	\textit{Science}.314(5806),1757–1761 (2006).
	
	\bibitem[2]{Hasan2010}M. Z. Hasan and C. L. Kane,
	\textit{Colloquium: Topological Insulators},
	\textit{Rev.Mod.Phys}.82,3045 (2010).
	
	\bibitem[3]{Fu2007}L. Fu, C. L. Kane, and E. J. Mele,
	\textit{Topological Insulators in Three Dimensions},
	\textit{Phys.Rev.Lett}.98,106803 (2007).
	
	\bibitem[4]{Shafiei2023}M. Shafiei, F.Fazileh, F.M.Peeters, M.V.Miloševic,
	\textit{High Chern number in strained thin films of dilute magnetic topological insulators},
	\textit{Phys.Rev.B}.107,195119(2023).
	
	\bibitem[5]{Haldane1988}F. D. M. Haldane, 
	\textit{Model for a Quantum Hall Effect without Landau Levels: Condensed-Matter Realization of the Parity Anomaly}, 
	\textit{Phys.Rev.Lett}.61,2015 (1988).
	
	\bibitem[6]{Rui2010}Yu Rui, Zhang W, Zhang HJ, Zhang SC, Dai X, Fang Z.
	\textit{Quantized anomalous Hall effect in magnetic topological insulators}, 
	\textit{SCIENCE}2;329(5987):61-4(2010).
	
	\bibitem[7]{Zhong2017}.Peichen Zhong, Yafei Ren, Yulei Han, Liyuan Zhang, and Zhenhua Qiao.
	\textit{In-plane magnetization-induced quantum anomalous Hall effect in atomic crystals of group-V elements}
	\textit{Phys. Rev. B }96, 241103 (2017).
	
	\bibitem[8]{Liu2011}Cheng-Cheng Liu,Wanxiang Feng,and Yugui Yao.
	\textit{Quantum Spin Hall Effect in Silicene and Two-Dimensional Germanium},
	\textit{Phys. Rev. Lett} 107, 076802 (2011).  
	
	\bibitem[9]{Xu2013} Y. Xu, B. Yan, H.-J. Zhang et al., 
	\textit{Large-Gap Quantum Spin Hall Insulators in Tin Films}
	\textit{Phys. Rev. Lett}. 111, 136804 (2013). 
	
	\bibitem[10]{Ezawa2015} M. Ezawa, 
	\textit{Monolayer Topological Insulators: Silicene, Germanene, and Stanene}, 
	\textit{J. Phys. Soc. Jpn}. 84, 121003 (2015).  
	
	\bibitem[11]{Cong2020} L. Cong, R. Singh,
	\textit{Spatiotemporal Dielectric Metasurfaces for Unidirectional Light Propagation},
	\textit{Adv.Mater}32(28):e2001418 (2020). 
	
	\bibitem[12]{Bentaibi2024PbBi}
	B. Bentaibi, L. B. Drissi, E. H. Saidi, O. Fassi-Fehri, M. Bousmina,
	\textit{Exploring topological phases in 2D half‑hydrogenated PbBi materials},
	\textit{Materials Science in Semiconductor Processing} 174,108180 (2024).
	
	\bibitem[13]{Bosnar2023}Bosnar, M., Vyazovskaya, A.Y., Petrov, E.K. et al.
	\textit{High Chern number van der Waals magnetic topological multilayers MnBi2Te4/hBN},
	\textit{ npj 2D Mater Appl} 7, 33 (2023).
	
	\bibitem[14]{Bentaibi2022OsC}
	B. Bentaibi, L. B. Drissi, E. H. Saidi, M. Bousmina, 
	\textit{New room‑temperature 2D hexagonal topological insulator OsC: First principles calculations},
	\textit{Materials Science in Semiconductor Processing}151,107009 (2022).
	
	\bibitem[15]{Qiao2010}Z.Qiao,S.A.Yang, W.Feng, W.K.Tse,J.Ding,Y.Yao,J.Wang, Q. Niu,
	\textit{Quantum Anomalous Hall Effect in Graphene from Rashba and Exchange Effects}, 
	\textit{Phys.Rev.B}.82,161414(R) (2010).
	
	\bibitem[16]{Zou2019}Zou, J., He, Z. and Xu, G.
	\textit{The study of magnetic topological semimetals by first principles calculations}. 
	\textit{npj Comput Mater} 5, 96 (2019).
	
	\bibitem[17]{Chen2020}Guo-Hong Chen,Yi-Nuo Chen, Yan-Wei Zhou,Yun-Lei Sun, and En-Jia Ye.
	\textit{Strain and electric field tunable electronic transport in armchair phosphorene nanodevice with normal-metal electrodes},
	\textit{AIP Advances} 10, 105012 (2020).
	
	\bibitem[18]{Liu2018}Hang Liu,Jia-Tao Sun,Cai Cheng, Feng Liu,Sheng Meng.
	\textit{Photoinduced Nonequilibrium Topological States in Strained Black Phosphorus},
	\textit{Phys. Rev. Lett}. 120, 237403 (2018).
	
	\bibitem[19]{Drissi2020} L.B.Drissi and E.H.Saidi,
	\textit{A signature index for third‑order topological insulators }
	\textit{J.Phys: Condensed Matter}.32,365701 (2020).
	
	\bibitem[20]{Benalcazar2017} W.A.Benalcazar, B.A.Bernevig, and T.L.Hughes,
	\textit{Quantized electric multipole insulators},
	\textit{Science} 357,61-66 (2017).
	
	\bibitem[21]{Schindler2018}Frank Schindler, et al.,
	\textit{Higher-order topological insulators},
	\textit{Science Advances},4(6). (2018).
	
	\bibitem[22]{Drissi2021}L. B. Drissi and E. H. Saidi,
	\textit{Domain walls in topological tri‑hinge matter},
	\textit{Eur.Phys.J.Plus}.136,68 (2021).
	
	\bibitem[23]{Park2019}Moon Jip Park,Youngkuk Kim.Gil Young Cho,and SungBin Lee,
	\textit{Higher-Order Topological Insulator in Twisted Bilayer Graphene},
	\textit{Phys. Rev. Lett}.123, 216803 (2021).
	
	\bibitem[24]{Drissi2022}L. B. Drissi, S. Lounis, E. H. Saidi, 
	\textit{Higher‑order topological matter and fractional chiral states},
	\textit{Eur.Phys.J.Plus} 137,796 (2022).
	
	\bibitem[25]{Noguchi2021}Ryo Noguchi, Masaru Kobayashi et al,
	\textit{Evidence for a higher-order topological insulator in a three-dimensional material},
	\textit{Nature Materials} 20,473–479 (2021).
	
	\bibitem[26]{Xue2019} Haoran Xue, Yahui Yang, Fei Gao, Yidong Chong, and Baile Zhang
	\textit{Acoustic higher-order topological insulator on a kagome lattice},
	\textit{Nature Materials volume} 18, pages108–112 (2019).
	
	\bibitem[27]{Benkaida2025}Benkaida, J., Benhaida, O., Drissi, L. B., and, Saidi, E. H.  
	\textit{Engineering Topological Transitions in Multi-Orbital Buckled Honeycomb Lattices.}
	\textit{Physics Letters A} \textbf{480}, 130954 (2025).  
	
	\bibitem[28]{St-Jean2017}St-Jean, P.Goblot, V.Galopin, E.Lemaître, A.Ozawa, T.Le Gratiet, L.Sagnes, I. Bloch, J., and, Amo, A.
	\textit{Lasing in topological edge states of a one-dimensional lattice.}
	\textit{Nature Photonics} \textbf{11}, 651–656 (2017).  
	
	\bibitem[29]{Gao2023}Gao, F., Xiang, X., Peng, Y.-G., et al.
	\textit{Orbital topological edge states and phase transitions in acoustic resonator chains.}
	\textit{Nature Communications} \textbf{14}, 8162 (2023).  
	
	\bibitem[30]{Sun2020} Gaoyong Sun, Wen-Long You , and,Tao Zhou.
	\textit{Topological phases of spinless p-orbital fermions in zigzag optical lattices},
	\textit{PHYSICAL REVIEW A},102, 063301 (2020).  
	
	\bibitem[31]{Barbosa2024}Barbosa, A. L. R., Rodrigues, F. M., and, Peres, N. M. R.
	\textit{Orbital Hall effect and topology on a two-dimensional triangular lattice: from bulk to edge.},
	\textit{Physical Review - Section B - Condensed Matter} 110: 085412 (2024).
	
	\bibitem[32]{Fan2023} Feng-Ren Fan, Cong Xiao, and Wang Yao, 
	\textit{Orbital Chern Insulator at $\nu=-2$ in Twisted MoTe$_{2}$},
	\textit{Physical Review B} 109.L041403  (2024).
	
	\bibitem[33]{Yao2026} Yueh-Ting Yao,Chia-Hung Chu,Arun Bansil, Hsin Lin, and Tay-Rong Chang.
	\textit{Orbital topology induced orbital Hall effect in two-dimensional insulators},
	\textit{Rep. Prog. Phys}. 89 018001 (2026) 
	
	\bibitem[34]{Lu2024}Lu, X., Jiang, R., and, Liu, J.
	\textit{Orbital magnetoelectric coupling of three-dimensional Chern insulators.} ,
	\textit{npj Quantum Materials} volume 10, Article number: 76 (2025).
	
	\bibitem[35]{Ezawa2012}Ezawa, M.
	\textit{Valley-polarized metals and quantum anomalous Hall effect in silicene.},
	\textit{New Journal of Physics}.14,033003 (2012).  
	
	\bibitem[36]{Won2020}Lee, K. W., and, Lee, C. E.
	\textit{Tunable valley polarization and anomalous Hall effect in bilayer MoTe$_2$.},
	\textit{Scientific Reports}.10, 11300 (2020).  
	
	\bibitem[37]{Slater1954} J.C.Slater and G. F. Koster,
	\textit{Simplified LCAO Method for the Periodic Potential Problem}, 
	\textit{Phys. Rev}.94,1498 (1954).
	
	\bibitem[38]{Reis2017} F.Reis et al,
	\textit{Bismuthene on a SiC substrate: A candidate for a high-temperature quantum spin Hall material},
	\textit{Science} 357,287290 (2017)
	
	\bibitem[39]{Song2018} Shi-Ru Song et al.,
	\textit{Dirac states from px,y orbitals in the buckled honeycomb structures: A tight-binding model and first-principles combined study},
	\textit{Chinese Phys.B}.27.087101 (2018).
	
	\bibitem[40]{Gardenier2020} T. S. Gardenier, J. J. van den Broeke, J. R.  Moes, I. Swart, C. Delerue, M. R. Slot, C. Morais ,Smith, and D.  Vanmaekelbergh,\textit{p-Orbital Flat Band and Dirac Cone in the Electronic Honeycomb Lattice},
	\textit{ACS Nano}14(10),13638-13644(2020).
	
	\bibitem[41]{Wan2023} Wenhui Wan et al, 
	\textit{Two-dimensional XY ferromagnetism above room temperature in Janus monolayer $V_{2}XN$ (X= P,As)},
	\textit{  Phys. Chem. Chem. Phys.},25,9311-9319(2023).
	
	\bibitem[42]{Altland1997} A. Altland and M. R. Zirnbauer,
	\textit{Nonstandard symmetry classes in mesoscopic normal-superconducting hybrid structures},
	\textit{Physical Review B} 55,1142 (1997)
	
	\bibitem[43]{Ryu2010} Shinsei Ryu et al, 
	\textit{Topological insulators and superconductors: tenfold way and dimensional hierarchy}, 
	\textit{New J. Phys}. 12 065010 (2010)
	
	\bibitem[44]{Beugeling2015}W. Beugeling, E. Kalesaki, C. Delerue,Y.-M. Niquet, D. Vanmaekelbergh and C. Morais Smith
	\textit{Topological states in multi-orbital HgTe honeycomb lattices},
	\textit{Nature Communications volume},6,: 6316 (2015).
	
	\bibitem[45]{Benhaida2025a} O. Benhaida, E. H. Saidi,L. B. Drissi, and  R.Ahl Laamara,
	\textit{Topological Properties of Bilayer $\alpha-T_{3}$ Lattice Induced by Polarized Light},
	\textit{ Advanced qute}.202500064 (2025)
	
	\bibitem[46]{Tamang2024}Lakpa Tamang,Sonu Verma, and Tutul Biswas.
	\textit{Orbital magnetization senses the topological phase transition in a spin-orbit coupled $\alpha-T_{3}$ systems.}  
	\textit{Phys.Rev.B},110,165426 (2024).
	
	\bibitem[47]{Benhaida2025b}O.Benhaida, E.H.Saidi, and L.B.Drissi,
	\textit{Optically controlled topological phases in the deformed  $\alpha-T_{3}$ lattice}, 
	\textit{Annals of Physics} 482:170203 (2025).
	
	\bibitem[48]{Benhaida2025c}O. Benhaida, L. B. Drissi, and E. H. Saidi,
	\textit{Extended Haldane Model in The Dice Lattice: Multiple Flat-Band-Induced Topological Transitions Revealed},
	\textit{Annalen der Physik} 538:2500615  (2026).
	
	\bibitem[49]{Li2024} Li, Y., Xu, T., and Zhang, D.
	\textit{Higher‑orbital Chern phases in photonic lattices with engineered coupling},
	\textit{Physical Review Applied} 20,014023 (2024).
	
	\bibitem[50]{Thouless1982} D. J. Thouless, M. Kohmoto, M. P. Nightingale, and M. den Nijs,
	\textit{Quantized Hall Conductance in a Two-Dimensional Periodic Potential},
	\textit{Phys. Rev. Lett}. 49, 405(1982).
	
	\bibitem[51]{Jungwirth2002} T.Jungwirth, Qian Niu, A.H.MacDonald.,
	\textit{Anomalous Hall Effect in Ferromagnetic Semiconductors},
	\textit{PhysRevLett}.88.207208(2002).
	
	\bibitem[52]{Xiao2010} Di Xiao. Ming-Che Chang. and Qian Niu.
	\textit{Berry phase effects on electronic properties},
	\textit{Rev. Mod. Phys}. 82, 1959 (2010) 
	
	\bibitem[53]{Sun2021} Shuo Sun; et al,
	\textit{Epitaxial Growth of Ultraflat Bismuthene with Large Topological Band Inversion Enabled by Substrate-Orbital-Filtering Effect}, 
	\textit{ACS Nano} 16,1436,(2022)
	
	\bibitem[54]{Deng2018} Deng, J. et al, 
	\textit{Epitaxial Growth of Ultraflat Stanene with Topological Band Inversion},
	\textit{Nat. Mater}.  17, 1081–1086 (2018).
	
	\bibitem[55]{Ren2018} Ren CC, Zhang SF, Ji WX, Zhang CW, Li P, Wang PJ. 
	\textit{Tunable Electronic and Topological Properties of Germanene by Functional Group Modification}.
	Nanomaterials (Basel). 6;8(3):145 (2018).
	
	\bibitem[56]{Qu2016} Jinfeng Qu, Xiangyang Peng, Di Xiao, and Jianxin Zhong,
	\textit{Giant spin splitting, strong valley selective circular dichroism and valley-spin coupling induced in silicene},
	\textit{Phys.Rev.B} 94, 075418 (2016)
	
	\bibitem[57]{Hattori2017} Ayami Hattori et al, 
	\textit{Edge states of hydrogen terminated monolayer materials: silicene, germanene and stanene ribbons},
	\textit{J.Phys.Condens.Matter} 29 115302 (2017)
	
\end{thebibliography}
\end{document}